\documentclass[10pt,a4paper,twoside]{article}    
\usepackage[T2A]{fontenc}
\usepackage[cp1251]{inputenc}
\usepackage[english,russian]{babel}
\usepackage{amssymb,amsmath,amsfonts}
\usepackage{stfi}
\usepackage{graphicx,epstopdf}
\newcommand{\issueYear}{February 2019}
\newcommand{\issueVolume}{Volume 61, Issue 10}

\begin{document}


\UDK{530.12:531.51:519.711.3}
\PACS{...}
  
\Title%
  {Dynamics of cosmological models with nonlinear classical phantom scalar fields. I. Formulation of the mathematical model}%
  {}

\Abstract%
  {Mathematical models describing the cosmological evolution of classical and phantom scalar fields with self-action are formulated and analyzed. Systems of dynamical equations in the plane, describing homogeneous cosmological models, have been obtained.  It is shown that depending on the parameters of the field model, it is possible to violate the singly-connectedness of the phase space of the corresponding dynamical model. }%
  {}

\Key%
  {phantom scalar field, quality analysis, asymptotic behavior, numerical simulation, numerical gravitation}%
  {}

\Datereceive{July 2, 2018.} 

\Author%
  {Yu.\,G.~Ignat'ev}
  {\textbf{Ignat'ev Yurii Gennadievich}, Doctor of Physics and Mathematics, Professor, Lobachevsky Institute of Mathematics and Mechanics, Kazan Federal University, ul. Kremlyovskaya, 35, Kazan, 420008, Russia.}
  {ignatev-yurii@mail.ru}  
  {}
  {}  

\Author%
  {A.\,A.~Agathonov}
  {\textbf{Agathonov Alexander Alexeevich}, Candidate of Physics and Mathematics, Assistant Professor, Lobachevsky Institute of Mathematics and Mechanics, Kazan Federal University, ul. Kremlyovskaya, 35, Kazan, 420008, Russia.}  
  {a.a.agathonov@gmail.com}
  {}
  {}

\begin{otherlanguage}{english}
\Header

\section{INTRODUCTION }

Standard cosmological models (SCMs) [1] based on a classical scalar field were investigated by methods of qualitative analysis of dynamical systems in [2–6, 8].  In [4], using methods of the qualitative theory of dynamical systems, a two-component cosmological model with minimal interaction was also investigated (see also [7, 8]).  In [9], a qualitative analysis was performed starting from scratch, and also a numerical analysis of a standard cosmological model based on a classical standard field, by reducing the problem to an investigation of a dynamical system in the two-dimensional phase plane $\{\Phi ,\dot \Phi\} $.  This study demonstrated the microscopic oscillatory character of the invariant cosmological acceleration at later stages of the expansion.  These results were then generalized to cosmological models with a $\lambda$-term [10–12], which confirmed  preservation of the oscillatory character of the invariant cosmological acceleration for quite small values of the cosmological term.  In addition to this, in these latter works, the possibility of a transition of the macroscopic acceleration to the nonrelativistic regime at late stages of the early Universe was demonstrated.  In still more recent works [13, 14], using direct averaging of the scalar potential and the derivative of the classical field over the microscopic oscillations, an effective macroscopic equation of state was obtained and it was shown that it tends to a nonrelativistic limit.  Moreover, in these works the macroscopic average of the square fluctuations of the scalar potential was calculated and it was shown that in late stages of the cosmological evolution, the mean-square energy of the microscopic oscillations of the scalar field exceeds by many orders of magnitude the energy of the macroscopic scalar field.  In [15] an approach proposed in the above-indicated works was implemented and applied to the two-component system \textit{scalar field + fluid} with arbitrary potential function $V(\phi)$.  

From a formal point of view, phantom fields were introduced in an obvious way into gravitation as one of several possible models of the scalar field in 1983 in [16].  In the indicated work, and also in later works (see, for example, [17, 18]), phantom fields were classified as scalar fields with attraction of like charged particles and distinguished by the factor $\varepsilon  =  - 1$ in the energy-momentum tensor of the scalar field. Note that phantom fields applied to wormholes and so-called black Universes were considered in [19, 20].  In the present work, in line with the generally accepted terminology, we use the term “phantom fields” to refer to scalar fields with a negative kinetic term in the energy-momentum tensor regardless of the sign of the potential term.  Here the negative potential term in the energy-momentum tensor corresponds to phantom scalar fields with attraction of like charged particles, and the positive potential term corresponds to phantom fields with repulsion.  In the first case, the signs of the kinetic and potential terms coincide, in the second case they are opposite, which is equivalent to changing the sign of the mass term in the Klein–Gordon equation.  The corresponding solutions for a solitary scalar charge do not take the form of a Yukawa potential, but rather the form of solutions of the equations of scalar Lipschitz perturbations for spherical symmetry ($\sin kr/r$) [21]. 

In subsequent works, a nonminimal theory of scalar interaction based on the concept of a fundamental scalar charge was successively developed both for classical and for phantom scalar fields [22–24].  In particular, in these works some peculiarities of phantom fields were clarified, for example, peculiarities of the interparticle interaction.  Later, these studies were deepened to extend the theory of scalar, including phantom, fields to the sector of negative particle masses, degenerate Fermi systems, conformally invariant interactions, etc. [25–29].  Mathematical models of scalar fields constructed in this way were applied to investigate the cosmological evolution of systems of interacting particles and scalar fields of both classical and phantom type [30– 32]. These studies clarified the unique peculiarities of the cosmological evolution of plasma with an interparticle phantom scalar interaction, such as the existence of giant bursts of cosmological acceleration, the presence of a plateau with constant acceleration, and other anomalies, sharply distinguishing the behavior of cosmological models with a phantom scalar field from models with a classical scalar field.  In particular, in  [32–34] a classification of types of behavior of cosmological models with an interparticle phantom scalar field was performed and four fundamentally different models were distinguished. These same works also pointed to the possibility of Bose condensation of nonrelativistic scalar-charged fermions under conditions of strong growth of the scalar field potential along with the possibility of considering such a condensate as a component of dark matter. It is important to note that in the case of a phantom field with attraction, over the course of cosmic evolution values of the cosmological acceleration greater than 1 were attainable, which corresponds, according to the generally accepted classification, to precisely a phantom state of matter.  

The indicated studies demonstrate the need to investigate both classical and phantom scalar fields with self-action as a possible basis of a cosmological model of the early Universe.  A preliminary qualitative analysis of a cosmological model based on a phantom scalar field with self-action was performed in [37–39].  In the present work, we analyze and lay out in detail the results of studies of cosmological models based on classical and phantom scalar fields.  In contrast to [25–39], we do not take into account the contribution of ordinary matter, i.e., we consider only free classical and phantom fields without a source.   

\section{MAIN EQUATIONS OF A COSMOLOGICAL MODEL BASED ON A SCALAR FIELD}

\subsection{Field equations}

The Lagrange function of the scalar field $\Phi$ with self-action has the form 
\begin{equation}
L=\frac{1}{8\pi }\left( {{e}_{1}}{{g}^{ik}}{{\Phi }_{,i}}{{\Phi }_{,k}}-2V(\Phi ) \right)
\end{equation}
Here  
\begin{equation}
V(\Phi )=-\frac{\alpha }{4}{{\left( {{\Phi }^{2}}-{{e}_{2}}\frac{{{m}^{2}}}{\alpha } \right)}^{2}} 
\end{equation}
where $\alpha$ is the self-action constant, $m$ is the mass of the quanta of the scalar field, and ${e_1} = 1$ for the classical scalar field and ${e_1} =  - 1$ for the phantom scalar field.  The potential energy $U$ is defined as      
\begin{equation}
4\pi U = V(\Phi ) \equiv  - \alpha \frac{{{\Phi ^4}}}{4} + {e_2}\frac{{{m^2}{\Phi ^2}}}{2} - \frac{{{m^4}}}{{4\alpha }} \Rightarrow 4\pi U =  - \alpha \frac{{{\Phi ^4}}}{4} + {e_2}\frac{{{m^2}{\Phi ^2}}}{2} + {\text{const}},
\end{equation}
so that    
\begin{equation}
\begin{gathered}
  U( - \alpha , - {e_2}, \pm \Phi ) \equiv  - U(\alpha ,{e_2}, \pm \Phi ) \hfill \\ 
\end{gathered}.
\end{equation}

The additive constant in the potential function can be discarded; therefore, in the limit of a small self-action constant we have a massive term in the Lagrangian (Eq. (1)): 
\begin{equation*}
U = {e_2}\frac{{{m^2}{\Phi ^2}}}{{4\pi }},\quad \alpha  \to 0.
\end{equation*}

Thus, we can conditionally distinguish the following cases: 
\begin{itemize}
\item
${e_2} = 1$ for a scalar field with attraction;  

\item
${e_2} = 1$ for a scalar field with repulsion. 
\end{itemize}

\begin{figure}[ht!]
\centering
\includegraphics[width=.6\textwidth]{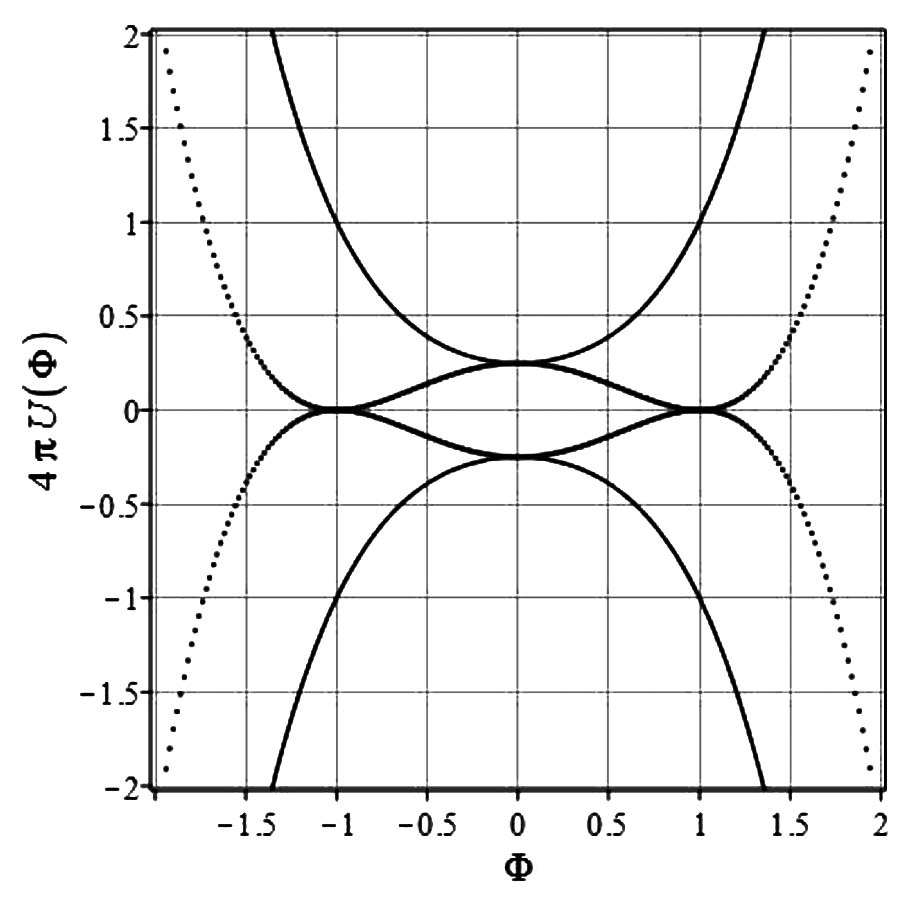}
\caption{Graphs of the potential energy $4\pi U(\alpha ,{e_2},\Phi )$. Parabolas (dotted curves): the upper curve corresponds to $4\pi U( - 1,1,\Phi )$ and the lower curve corresponds to $4\pi U(1, - 1,\Phi )$,  Parabolas with three extrema (solid curves): the upper curve corresponds to $4\pi U( - 1, - 1,x)$ and the lower curve corresponds to  $4\pi U(1,1,x)$.  The upper parabolas correspond to attraction while the lower ones correspond to repulsion.}
\label{img:01}
\end{figure}   

The energy-momentum tensor of the scalar field for the Lagrange function given by Eq. (1) takes the following standard form: 
\begin{equation}
{T_{ik}} = \frac{1}{{8\pi }}(2{e_1}{\Phi _{,i}}{\Phi _{,k}} - {g_{ik}}{\Phi _{,j}}{\Phi ^{,j}} + 2V(\Phi ){g_{ik}}).
\end{equation} 

Setting the covariant divergence of this tensor equal to zero leads to the equation of the free scalar field:  
\begin{equation}
\square \Phi  + V'(\Phi ) = 0.
\end{equation}

Since it is possible to add an arbitrary constant to the Lagrange function, in what follows we will omit the corresponding constant in the potential function in those places where this leads to simplifications.  Such a renormalization returns us to the original Lagrange function of the scalar field with self-action of [37–39], which we will also make use of (see  [16]): 
\begin{equation}
L = \frac{1}{{8\pi }}\left( {{e_1}{g^{ik}}{\Phi _{,i}}{\Phi _{,k}} - {e_2}{m^2}{\Phi ^2} + \frac{\alpha }{2}{\Phi ^4}} \right).
\end{equation}

The energy-momentum tensor for the Lagrangian given by Eq. (7) is equal to 
\begin{equation}
{{T_{ik}} = \frac{1}{{8\pi }}\left( {2{e_1}{\Phi _{,i}}{\Phi _{,k}} - {g_{ik}}{e_1}{\Phi _{,j}}{\Phi ^{,j}} + {g_{ik}}{e_2}{m^2}{\Phi ^2} - {g_{ik}}\frac{\alpha }{2}{\Phi ^4}} \right)}.
\end{equation}

Using the Lagrange function in the form given by Eq. (7), we obtain from Eq. (6) the following equation: 
\begin{equation}
\square \Phi  + m_*^2\Phi  = 0,
\end{equation}
where ${m_*}$ is the effective mass of a scalar boson
\begin{equation}
m_*^2 \equiv {e_1}({e_2}{m^2} - \alpha {\Phi ^2}),
\end{equation}
which, in principle, can also be an imaginary quantity.    

We also write out the Einstein equations with the cosmological term ($G = c = \hbar  = 1$; the Ricci tensor is obtained by convolution of the first and fourth indices  ${R_{ik}} = R_{ ikj}^j$; the metric has the signature $( - 1, - 1, - 1, + 1)$) 
\begin{equation}
{R^{ik}} - \frac{1}{2}R{g^{ik}} = \lambda {g^{ik}} + 8\pi {T^{ik}},
\end{equation}
where $\lambda  \geqslant 0$ is the cosmological constant.

\subsection{Equations of the cosmological model}

We write out the self-consistent system of equations of the cosmological model based on a free scalar field and the spatially-flat Friedmann metric (Eqs. (9)–(11)), setting $\Phi  = \Phi (t)$: 
\begin{equation}
d{s^2} = d{t^2} - {a^2}(t)(d{x^2} + d{y^2} + d{z^2}).
\end{equation}

The indicated system consists of one Einstein equation 
\begin{equation}3\frac{{{{\dot a}^2}}}{{{a^2}}} = {e_1}{\dot \Phi ^2} + {e_2}{m^2}{\Phi ^2} - \frac{\alpha }{2}{\Phi ^4} + \lambda \end{equation}
and the equation of the scalar field  
\begin{equation}\ddot \Phi  + 3\frac{{\dot a}}{a}\dot \Phi  + m_*^2\Phi  = 0,
\end{equation}
where $\dot f \equiv df/dt$.  Here the energy-momentum tensor (Eq. (8)) has the structure of the energy-momentum tensor of an isotropic fluid with energy density and pressure 
\begin{equation}
\begin{gathered}
  \varepsilon  = \frac{1}{{8\pi }}\left( {{e_1}{{\dot \Phi }^2} + {e_2}{m^2}{\Phi ^2} - \frac{\alpha }{2}{\Phi ^4}} \right), \hfill \\
  p = \frac{1}{{8\pi }}\left( {{e_1}{{\dot \Phi }^2} - {e_2}{m^2}{\Phi ^2} + \frac{\alpha }{2}{\Phi ^4}} \right), \hfill \\ 
\end{gathered}
\end{equation}
so that    
\begin{equation*}\varepsilon  + p = \frac{{{e_1}}}{{4\pi }}{\dot \Phi ^2}.
\end{equation*}

In what follows, we will also have need of the values of the two kinematic functions of the Friedmann Universe: 
\begin{equation}
H(t) = \frac{{\dot a}}{a} \geqslant 0,\,\,\,\;\;\Omega (t) = \frac{{a\ddot a}}{{{{\dot a}^2}}} \equiv 1 + \frac{{\dot H}}{{{H^2}}}
\end{equation}
-- the Hubble constant $H(t)$ and the invariant cosmological acceleration $Omega (t)$, which is an invariant and is expressed in the following way with the help of the \textit{barotropic coefficient} $\varkappa =p/\varepsilon$: 
\begin{equation}
\Omega =-\frac{1}{2}(1+3\varkappa ).
\end{equation}
Differentiating Eq. (13) with definition (16) and the field equation (Eq. (14)) taken into account, we find  
\begin{equation}
\dot H =  - {e_1}{\dot \Phi ^2}.
\end{equation}

Thus, for classical scalar fields $\dot H \leqslant 0$ and for phantom scalar fields $\dot H \geqslant 0$; therefore, for classical fields $Omega  \leqslant 1$, and for phantom fields $\Omega  \geqslant 1$.

Let us consider possible particular cases.  For a\textit{ classical scalar field with attraction} (${e_1} = {e_2} = 1$) system of equations (13) and (14) takes the following form:
\begin{equation}    
3{H^2} = {\dot \Phi ^2} + {m^2}{\Phi ^2} - \frac{\alpha }{2}{\Phi ^4} + \lambda,
\end{equation}
\begin{equation}
\ddot \Phi  + 3H\dot \Phi  + {m^2}\Phi  - \alpha {\Phi ^3} = 0,
\end{equation}
and the energy density and pressure of the scalar field are equal to    
\begin{equation}
\begin{gathered}
  \varepsilon  = \frac{1}{{8\pi }}\left( {{{\dot \Phi }^2} + {m^2}{\Phi ^2} - \frac{\alpha }{2}{\Phi ^4}} \right), \hfill \\
  p = \frac{1}{{8\pi }}\left( {{{\dot \Phi }^2} - {m^2}{\Phi ^2} + \frac{\alpha }{2}{\Phi ^4}} \right). \hfill \\ 
\end{gathered}
\end{equation}

For a \textit{classical scalar field with repulsion} (${e_1} = 1,\,\,\,\,{e_2} =  - 1$, $m_*^2 =  - {m^2} - \alpha {\Phi ^2}$) system of equations (13) and (14) takes the form  
\begin{equation}
3{H^2} = {\dot \Phi ^2} - {m^2}{\Phi ^2} - \frac{\alpha }{2}{\Phi ^4} + \lambda,
\end{equation}
\begin{equation}
\ddot \Phi  + 3H\dot \Phi  - {m^2}\Phi  - \alpha {\Phi ^3} = 0,
\end{equation}
and the energy density and pressure of the scalar field are equal to  
\begin{equation}
\begin{gathered}
  \varepsilon  = \frac{1}{{8\pi }}\left( {{{\dot \Phi }^2} - {m^2}{\Phi ^2} - \frac{\alpha }{2}{\Phi ^4}} \right), \hfill \\
  p = \frac{1}{{8\pi }}\left( {{{\dot \Phi }^2} + {m^2}{\Phi ^2} + \frac{\alpha }{2}{\Phi ^4}} \right). \hfill \\ 
\end{gathered}
\end{equation}

For a\textit{ phantom scalar field with attraction}  (${e_1} =  - 1,\,\,\,{e_2} = 1$, $m_*^2 =  - {m^2} - \alpha {\Phi ^2}$) 
\begin{equation}
3{H^2} =  - {\dot \Phi ^2} + {m^2}{\Phi ^2} - \frac{\alpha }{2}{\Phi ^4} + \lambda,
\end{equation}
\begin{equation}
\ddot \Phi  + 3H\dot \Phi  - {m^2}\Phi  + \alpha {\Phi ^3} = 0,
\end{equation}
\begin{equation}
\begin{gathered}
  \varepsilon  = \frac{1}{{8\pi }}\left( { - {{\dot \Phi }^2} + {m^2}{\Phi ^2} - \frac{\alpha }{2}{\Phi ^4}} \right), \hfill \\
  p = \frac{1}{{8\pi }}\left( { - {{\dot \Phi }^2} - {m^2}{\Phi ^2} + \frac{\alpha }{2}{\Phi ^4}} \right). \hfill \\ 
\end{gathered}
\end{equation}

For a \textit{phantom scalar field with repulsion}  (${e_1} = {e_2} =  - 1$, $m_*^2 = {m^2} - \alpha {\Phi ^2}$) 
\begin{equation}
3{H^2} =  - {\dot \Phi ^2} - {m^2}{\Phi ^2} - \frac{\alpha }{2}{\Phi ^4} + \lambda,
\end{equation}
\begin{equation}
\ddot \Phi  + 3H\dot \Phi  + {m^2}\Phi  + \alpha {\Phi ^3} = 0,
\end{equation}
\begin{equation}
\begin{gathered}
  \varepsilon  = \frac{1}{{8\pi }}\left( { - {{\dot \Phi }^2} - {m^2}{\Phi ^2} - \frac{\alpha }{2}{\Phi ^4}} \right), \hfill \\
  p = \frac{1}{{8\pi }}\left( { - {{\dot \Phi }^2} + {m^2}{\Phi ^2} + \frac{\alpha }{2}{\Phi ^4}} \right). \hfill \\ 
\end{gathered}
\end{equation}

\section{QUALITATIVE ANALYSIS}

\subsection{Reduction of the system of equations to normal form}

Noting that it follows from the Einstein equation (Eq. (13)) that the Hubble constant can be expressed in terms of the functions $\Phi \;\,{\text{and}}\,{\kern 1pt} \,\dot \Phi $, transforming to dimensionless Compton time 
\begin{equation}
mt=\tau \quad (m0),
\end{equation}
and making the standard substitution of variables $\Phi ' = Z(\tau )$ ($f' \equiv df/d\tau$), we reduce the Einstein equation (Eq. (13)) to dimensionless form: 
\begin{equation}
H'\;_m^2 = \frac{1}{3}\left[ {{e_1}{Z^2} + {e_2}{\Phi ^2} - \frac{{{\alpha _m}}}{2}{\Phi ^4} + {\lambda _m}} \right],
\end{equation}
and the field equation (Eq. (14)) to the form of a normal autonomous system of ordinary differential equations in the plane $\{ \Phi ,Z\}$: 
\begin{equation}
\begin{gathered}
  \Phi ' = Z, \\ 
  Z' =  - \sqrt 3 Z\sqrt {{e_1}{Z^2} + {e_2}{\Phi ^2} - \frac{{{\alpha _m}}}{2}{\Phi ^4} + {\lambda _m}}  - {e_1}{e_2}\Phi  + {e_1}{\alpha _m}{\Phi ^3}, \\ 
\end{gathered}
\end{equation}
where we have introduced the following notation:   
\begin{equation*}
{\lambda _m} \equiv \frac{\lambda }{{{m^2}}},\,\,\,\,\;{\alpha _m} \equiv \frac{\alpha }{{{m^2}}}.
\end{equation*}
Here   
\begin{equation}
\frac{{a'}}{a} \equiv \Lambda ' = {H_m} \equiv \frac{H}{m},\,\,\quad \Omega  = \frac{{aa''}}{{{{a'}^2}}} \equiv 1 + \frac{{h'}}{{{h^2}}}
\end{equation}
where    
\begin{equation}
\Lambda  = \ln a(\tau ).
\end{equation}

Note that in this notation all of the quantities  $\Phi ,\,\,Z,\,\,{H_m},\,\,{\alpha _m},\,\,\Omega ,\,\,{\text{and}}\;\tau $ are dimensionless, and time $\tau$ is measured in Compton units.  

Thus we have an autonomous two-dimensional dynamical system in the phase plane $\{ \Phi ,Z\}$. To reduce it to the standard notation of the qualitative theory of differential equations (see, for example, [15]), we set   
\begin{equation}
\begin{gathered}
  \Phi  = x,\,\,\,\,\;Z = y, \\ 
  P(x,y) = y, \\ 
  Q(x,y) =  - \sqrt 3 y\sqrt {{e_1}{y^2} + {e_2}{x^2} - \frac{{{\alpha _m}}}{2}{x^4} + {\lambda _m}}  - {e_1}{e_2}x + {e_1}{\alpha _m}{x^3}. \\ 
\end{gathered}
\end{equation}

The corresponding normal system of equations in the standard notation has the form  
\begin{equation}
x' = P(x,y),\,\;\;\;y' = Q(x,y).
\end{equation}

In order for system of differential equations (33) (or (37)) to have a real solution, it is necessary to satisfy the inequality (Figs. 2–4) 
\begin{equation}
{e_1}{y^2} + {e_2}{x^2} - \frac{{{\alpha _m}}}{2}{x^4} + {\lambda _m} \geqslant 0.
\end{equation}

The boundary of region (38) is defined by the fourth-order algebraic equation   
\begin{equation}
{e_1}{y^2} + {e_2}{x^2} - \frac{{{\alpha _m}}}{2}{x^4} + {\lambda _m} = 0.
\end{equation}

Introducing the variables   
\begin{equation}
\xi  = {x^2} - \frac{{{e_2}}}{{{\alpha _m}}}\,\,\,\,{\text{and}}\,\,\;\eta  = {y^2} + {e_1}\left( {\frac{1}{{2{\alpha _m}}} + \lambda } \right),
\end{equation}
it is possible to transform inequality (38) into the inequalities    
\begin{equation}
\begin{gathered}
  {\xi ^2} \leqslant \frac{{2{e_1}}}{{{\alpha _m}}}\eta ,\quad \alpha  > 0, \hfill \\
  {\xi ^2} \geqslant \frac{{2{e_1}}}{{{\alpha _m}}}\eta ,\quad \alpha  < 0, \hfill \\ 
\end{gathered}
\end{equation}
and Eq. (39) into the canonical equation of a parabola on the $\eta$ axis: 
\begin{equation}
{\xi ^2} = \frac{{2{e_1}}}{{{\alpha _m}}}\eta.
\end{equation}

Note that for a classical field (${e_1} = 1$) in the case $\alpha  > 0$ it is always the case that $\eta  > 0$; therefore, the variable  $\xi$ can only take values in a bounded interval. 

Note also that the fourth-order curve described by Eq. (39) corresponds to the zero value of the Hubble constant, i.e., zero expansion velocity.  On the other hand, if as the effective energy of matter we consider the sum of the energy of the scalar field and the contribution of the cosmological constant 
\begin{equation}
{{\varepsilon _{{\text{eff}}}} = \varepsilon  + \frac{1}{{8\pi }}\lambda  \equiv  = {m^2}{\varepsilon _m} = \frac{{{m^2}}}{{8\pi }}\left( {{e_1}{y^2} + {e_2}{x^2} - \frac{{{\alpha _m}}}{2}{x^4} + {\lambda _m}} \right)},
\end{equation}
then the curve described by Eq. (39) will correspond to zero effective energy ${\varepsilon _{{\text{eff}}}} = 0$. For $\lambda  = 0$ the outer region bounded by this curve is the region of accessible values of the dynamical variables for the classical field (Fig. 2a); for the phantom field, the accessible values of the dynamical variables lie, on the contrary, in the inner region (Fig. 2b).  For $\lambda 0$ the property of reversibility disappears and it is possible to divide the regions of accessible values of the dynamical variables into two symmetric regions (Figs. 3 and 4).  Thus, nonlinear scalar fields, both classical and phantom, are characterized by violation of singly-connectedness of phase space and the appearance of the white regions, which are inaccessible for dynamical systems.  This fact can lead to fundamental differences in the asymptotic behavior of cosmological models based on nonlinear scalar fields. 

The work was performed according to the Russian Government Program of Competitive Growth of Kazan Federal University. 

\begin{figure}[h!]
\centering
\includegraphics[width=.95\textwidth]{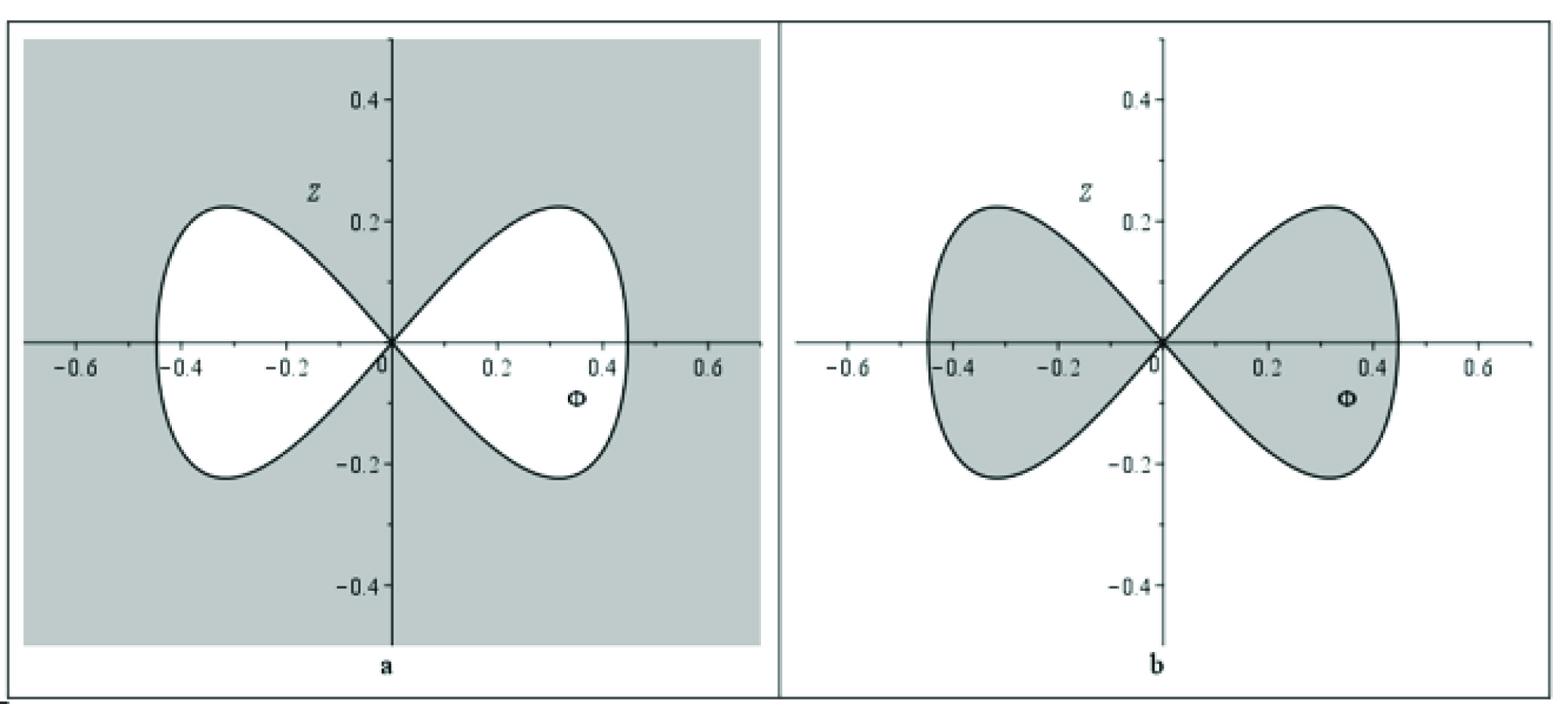}
\caption{Classical scalar field (${e_1} = 1,\;{e_2} =  - 1,\;\alpha  =  - 10,\lambda  = 0$) (a). Phantom scalar field (${e_1} =  - 1,\;{e_2} = 1,\;\alpha  = 10,\lambda  = 0$) (b).  Motion is possible only in the filled-in regions.}
\label{img:02}
\end{figure}

\begin{figure}[h!]
\centering
\includegraphics[width=.95\textwidth]{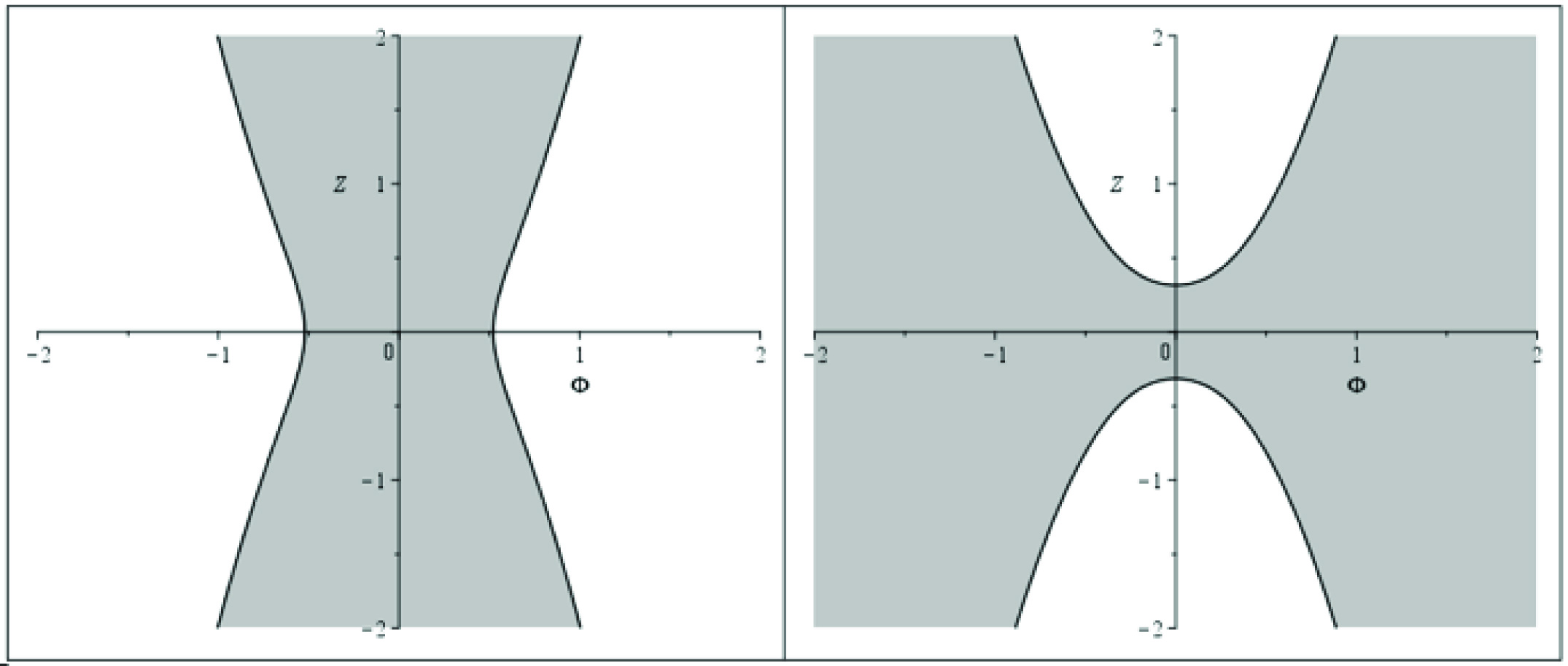}
\caption{Classical scalar field (${e_1} = 1,\;{e_2} = 1,\;\alpha  = 10,\lambda  = 0.1$) (a). Phantom scalar field  (${e_1} =  - 1,\;{e_2} = 1,\;\alpha  =  - 10,\lambda  = 0$) (b). Motion is possible only in the filled-in regions.}
\label{img:03}
\end{figure}

\begin{figure}[h!]
\centering
\includegraphics[width=.95\textwidth]{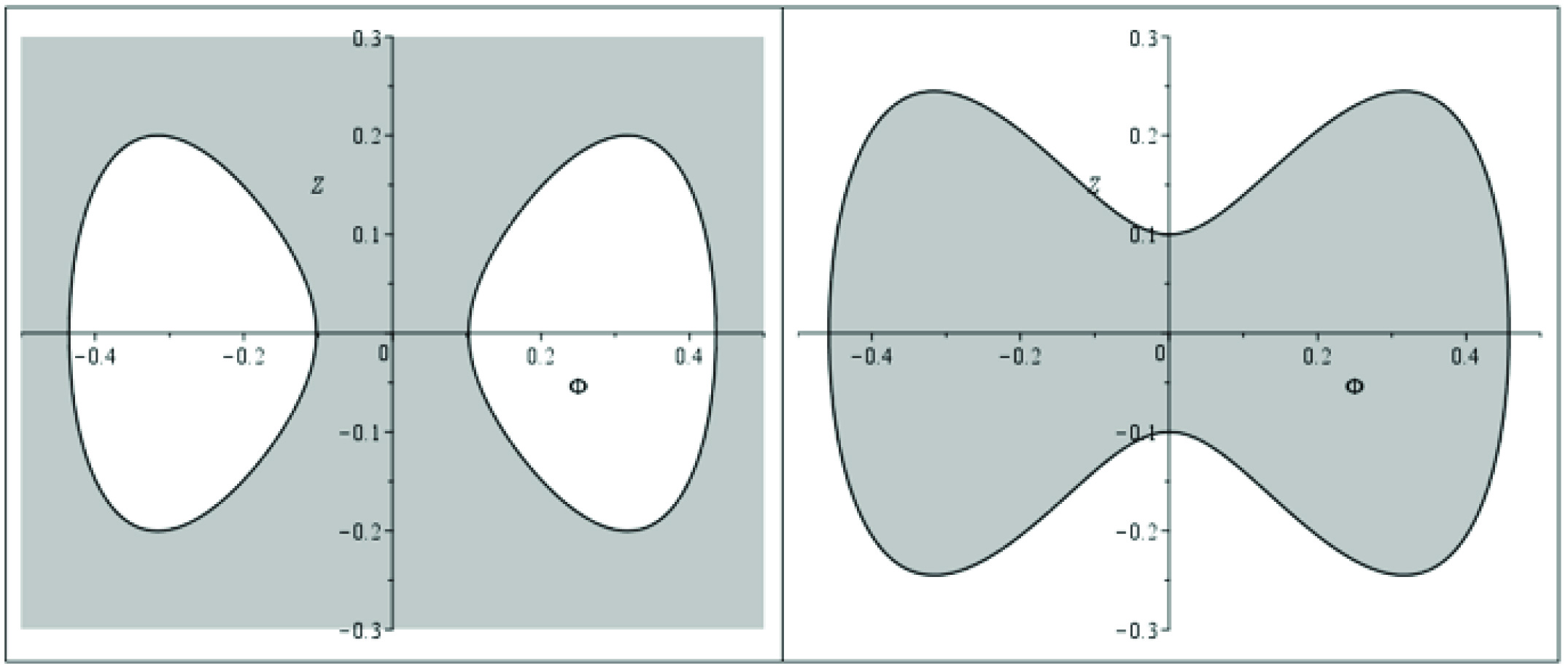}
\caption{Classical scalar field  (${e_1} = 1,\;{e_2} =  - 1,\;\alpha  =  - 10,\lambda  = 0.01$) (a). Phantom scalar field  (${e_1} =  - 1,\;{e_2} = 1,\;\alpha  = 10,\lambda  = 0.01$) (b). Motion is possible only in the filled-in regions.}
\label{img:04}
\end{figure}


\end{otherlanguage}

%


\begin{thebibliography}{99}


\bibitem{01}
D. S. Gorbunov and V. A. Rubakov, Introduction to the Theory of the Early Universe: Cosmological Perturbations and Inflationary Theory, World Scientific, Singapore (2011). 

\bibitem{02}
V. A. Belinskii, L. P. Grishchuk, Ya. B. Zel’dovich, and I. M. Khalatnikov, Zh. Eksp. Teor. Fiz., 89, 346–355 (1985).  

\bibitem{03}
A. D. Dolgov, Ya. B. Zel’dovich, and M. V. Sazhin,  Cosmology of the Early Universe [in Russian], Moscow State Univ. Press, Moscow (1988).  

\bibitem{04}
V. M. Zhuravlev, Zh. Eksp. Teor. Fiz., 120, No. 5, 1043–1061 (2001).  

\bibitem{05}
L. A. Urena-Lopez and M. J. Reyes-Ibarra, arXiv: 0709.3996v2 [astro-ph] (2009).

\bibitem{06}
K. A. Bronnikov and S. G. Rubin, Lectures on Gravitation and Cosmology [in Russian], National Research Nuclear Univ. Press, Moscow (2008).     

\bibitem{07}
V. M. Zhuravlev, T. V. Podymova, and E. A. Pereskokov, Grav. Cosmol., 17, No. 2, 101–109 (2011).

\bibitem{08}
L. A. Urena-Lopez, arXiv:1108.4712v2 [astro-ph.CO] (2012).

\bibitem{09}
Yu. G. Ignat’ev, Vremya i Fundamental’nye Vzaimodelistviya, No. 3, 16–36 (2016) [arXiv:1609.00745v1[gr-qc] (2016)].    

\bibitem{10}
Yu. G. Ignat’ev, Vremya i Fundamental’nye Vzaimodelistviya, No. 3, 37–47 (2016).

\bibitem{11}
Yu. G. Ignat'ev, Grav. Cosmol., No. 23, 131–141 (2017); arXiv:1609.00745 [gr-qc], arXiv:1609.08851 [gr-qc] (2016). 

\bibitem{12}
Yu. G. Ignat’ev, Classical Cosmology and Dark Energy [in Russian], Academy of Sciences of the Republic of Tatarstan, Kazan (2016).  

\bibitem{13}
Yu. G. Ignat'ev and A. R. Samigullina, Russ. Phys. J., 60, No. 7, 1173–1181 (2017).  

\bibitem{14}
Yu. G. Ignat'ev, D.Yu. Ignatyev, and A. R. Samigullina, Grav. Cosmol., 24, No 2, 148–153 (2018); arXiv:1705.05000 [gr-qc].

\bibitem{15}
V. M. Zhuravlev, Prostranstvo, Vremya i Fundamental’nye Vzaimodelistviya, No. 4, 39–51 (2016).    

\bibitem{16}
Yu. G. Ignat'ev, Russ. Phys. J., 26, No. 12, 1065–1067 (1983).

\bibitem{17}
Yu. G. Ignat'ev, Kuzeev R. R., Ukr. Fiz. Zh., 29, 1021 (1984).

\bibitem{18}
Yu. G. Ignatyev and R. F. Miftakhov, Grav. Cosmol., 12, No. 2–3, 179 (2006).

\bibitem{19}K. A. Bronnikov and J. C. Fabris, Phys. Rev. Lett., 96, 973 (2006).

\bibitem{20}
S. V. Bolokhov, K. A. Bronnikov, and M. V. Skvortsova, Class. Quantum Grav., No. 29, 245006 (2012).

\bibitem{21}
Yu. G. Ignat'ev, Russ. Phys. J.,  55, No. 11, 1345–1350 (2012).

\bibitem{22}
Yu. G. Ignat'ev, Russ. Phys. J.,  55, No. 2, 166–172 (2012).

\bibitem{23}
Yu. G. Ignat'ev, Russ. Phys. J.,  55, No. 5, 550–560 (2012).

\bibitem{24}
Yu. G. Ignat'ev, The Nonequilibrium Universe: Kinetic Models of Cosmological Evolution, Kazan University Press, Kazan (2013). 

\bibitem{25}
Yu. G. Ignat'ev, Vremya i Fundamental’nye Vzaimodelistviya, No. 1, 47–69 (2014).

\bibitem{26}
Yu. G. Ignatyev and D. Yu. Ignatyev, Grav. Cosmol., 20, No. 4, 299–303 (2014).

\bibitem{27}
Yu. G. Ignatyev, A. A. Agathonov, and D. Yu. Ignatyev, Grav. Cosmol., 20, No. 4, 304–308 (2014).

\bibitem{28}
Yu. G. Ignatyev, Grav. Cosmol., 21, No. 4, 296–308 (2015).

\bibitem{29}
Yu. G. Ignatyev and A. A. Agathonov, Grav. Cosmol., 21, No. 2, 105–112 (2015).

\bibitem{30}
Yu. G. Ignat'ev and M. L. Mikhailov, Russ. Phys. J., 57, No. 12, 1743–1752 (2015).   

\bibitem{31}
Yu. Ignat'ev, A. Agathonov, M. Mikhailov, and D. Ignatyev, Astr. Space Sci., 357, 61 (2015).

\bibitem{32}
Yu. G. Ignat'ev and A. A. Agathonov, Vremya i Fundamental’nye Vzaimodelistviya, No. 3, 48–90 (2016).  

\bibitem{33}
Yurii Ignat'ev, Alexander Agathonov, and Dmitry Ignatyev, arXiv:1608.05020 [gr-qc] (2016).

\bibitem{34}
Yu. G. Ignat'ev, A. A. Agathonov, and D. Yu. Ignatyev, Grav. Cosmol., 24, No. 1, 1–12 (2018).   

\bibitem{35}
Yu. G. Ignat'ev, Vremya i Fundamental’nye Vzaimodelistviya, No. 1, 79–98 (2012).  

\bibitem{36}
Yu. G. Ignat'ev and M. L. Mikhaikov, Vremya i Fundamental’nye Vzaimodelistviya, No. 4, 75–90 (2015).  

\bibitem{37}
Yu. G. Ignat'ev, Russ. Phys. J.,  59, No. 1, 20–31 (2016).    

\bibitem{38}
Yu. G. Ignat'ev and A. A. Agathonov, Vremya i Fundamental’nye Vzaimodelistviya, No. 4, 52–61 (2016).  

\bibitem{39}
Yu. G. Ignat'ev and A. A. Agathonov, Grav. Cosmol., 23, No. 3, 230–235 (2016) arXiv:1610.04443 [gr-qc] (2016). 


\end{thebibliography}
\end{document}